\documentclass[final,3p,times,twocolumn]{elsarticle}
 \biboptions{comma,sort&compress}
\usepackage{here}
 \usepackage{graphicx}
  \usepackage{epsfig,epstopdf}
\usepackage{amsmath}
\usepackage{amssymb}
\def\nin{\noindent}
\def\beq{\begin{equation}}
\def\eeq{\end{equation}}
\def\bea{\begin{eqnarray}}
\def\eea{\end{eqnarray}}

\def\la{\langle}
\def\ra{\rangle}
\def\GeV{{\rm GeV}}
\def\A{{\cal A}}
\def\W{{\cal W}}
\newcommand{\m}{\tilde{m}}
\newcommand{\barq}{\protect{\bar{q}}}
\newcommand{\barQ}{\protect{\bar{Q}}}
\newcommand{\wf}{\varphi}
\newcommand{\vr}{\vec{R}}
\newcommand{\Sp}{{\cal S}}
\newcommand{\Sps}{{\rm S}}

\newcommand{\DXtot}{[{\cal D}X^{(\wedge),(\vee)}]}
\newcommand{\DP}{[{\cal D}\Pi]}
 
\newcommand{\slapi}{{\raisebox{0.8pt}{\makebox[0pt][l]{\hspace{-0.2pt}\bf{
          /}}}{\Pi}}} 
\newcommand{\de}{\partial}
\newcommand{\f}[2]{\frac{#1}{#2}}

\journal{Nucl. Phys. (Proc. Suppl.)}

\begin{document}

\begin{frontmatter}

\title{Wilson-loop formalism for Reggeon exchange at high energy}
 \author{Matteo Giordano\corref{CPAN}}
 \address{Departamento de F\'isica Te\'orica, Universidad de
  Zaragoza, \\ 
  Calle Pedro Cerbuna 12, E--50009 Zaragoza, Spain}
 \cortext[CPAN]{Supported by MICINN [CPAN project CSD2007-00042
   (Consolider-Ingenio2010 program); grant FPA2009-09638]. 
}
\ead{giordano@unizar.es}

\begin{abstract}
\noindent
I will discuss how the non-vacuum, quark-antiquark Reggeon-exchange
contribution to meson-meson elastic scattering, at high energy and low
tranferred momentum, can be related to the path-integral of a certain
Wilson-loop expectation value over the trajectories of the exchanged
fermions. Making use of this representation, I will show how a linear
Regge trajectory is obtained through gauge/gravity duality and the use
of minimal surfaces.
\end{abstract}

\begin{keyword}
Nonperturbative effects\sep QCD \sep Gauge-gravity correspondence
\end{keyword}

\end{frontmatter}

\section{Introduction}
\nin
The study of soft high-energy scattering (SHES) in strong interactions
($\sqrt{s}\gg 1~\GeV$, $\sqrt{|t|}\lesssim 1~\GeV$) dates back to the
mid '50s, well before the discovery of QCD. One of the key concepts in
the phenomenological description of SHES 
 is that of Regge
poles, i.e., singularities in the complex-angular-momentum plane of
the $t$-channel amplitude, corresponding in physical terms to the
exchange of families of states between the colliding hadrons. The
position of these singularities varies with $t$ along the so-called
Regge trajectories $\alpha(t)$, and governs the high-energy behaviour
of the scattering amplitudes, i.e., $\A(s\!\to\!\infty,t)\sim
s^{\alpha(t)}$. 
The dominant trajectory in the elastic channel is called
{\it Pomeron}, and corresponds to the exchange of states with vacuum
quantum numbers, while the subleading non-vacuum trajectories are
usually called {\it Reggeons}. The explanation of these
concepts from first principles is an open problem in
QCD, involving its nonperturbative (NP), strong-coupling regime. 

A NP approach to SHES in the framework of QCD has been
proposed some time ago~\cite{Nac}. This approach adopts a partonic
description of hadrons over a small time-window at interaction time,
over which partons do not split or annihilate and can be treated as in
and out states of a scattering process. Starting from the corresponding
amplitudes, one then reconstructs the hadronic amplitudes by folding
with appropriate wave functions describing the hadrons. In the case of
the Pomeron exchange (PE) process, the partons travel approximately on
their classical trajectories, as the energy is large, and are
practically undisturbed by the diffusion process, as the momentum
transfer is small, and so can be treated in an eikonal
approximation~\cite{Nac,Dos,Meggiola}. In the case of Reggeon
exchange (RE) the picture involves the exchange of a pair of valence quark
and antiquark, and a different treatment is required. An approach
based on the path-integral (PI) representation for the fermion
propagator~\cite{Brandt} and on analytic continuation (AC) to Euclidean
space~\cite{Meggiolaro97,crossing,EMduality} has been suggested a 
few years ago~\cite{Jani}, but a complete derivation was lacking until
recently~\cite{reggeWL}.  

\section{Reggeon Exchange and Wilson Loops}
\nin
We briefly sketch now the derivation of a NP expression for the RE
contribution to SHES amplitudes. We focus on the elastic scattering of
two mesons $M_{1,2}$, of masses $m_{1,2}$, taken for simplicity with
the following flavour content, $M_1=Q\bar q$, $M_2=q\bar Q'$. In the
soft high-energy regime, the initial momenta $p_{1,2} =
m_{1,2}u_{1,2}$, with $u_i$ purely longitudinal, $u_i^2=1$ 
and $u_1\cdot u_2 = \cosh\chi$, are practically unchanged by the
scattering process, i.e., $p_i'\simeq p_i$, with the transferred
momentum $q =  p_1-p_1' \simeq  (0,0,\vec{q}_\perp)$. 

\begin{figure}[hbt] 
  \centering
    \includegraphics[width=0.23\textwidth]{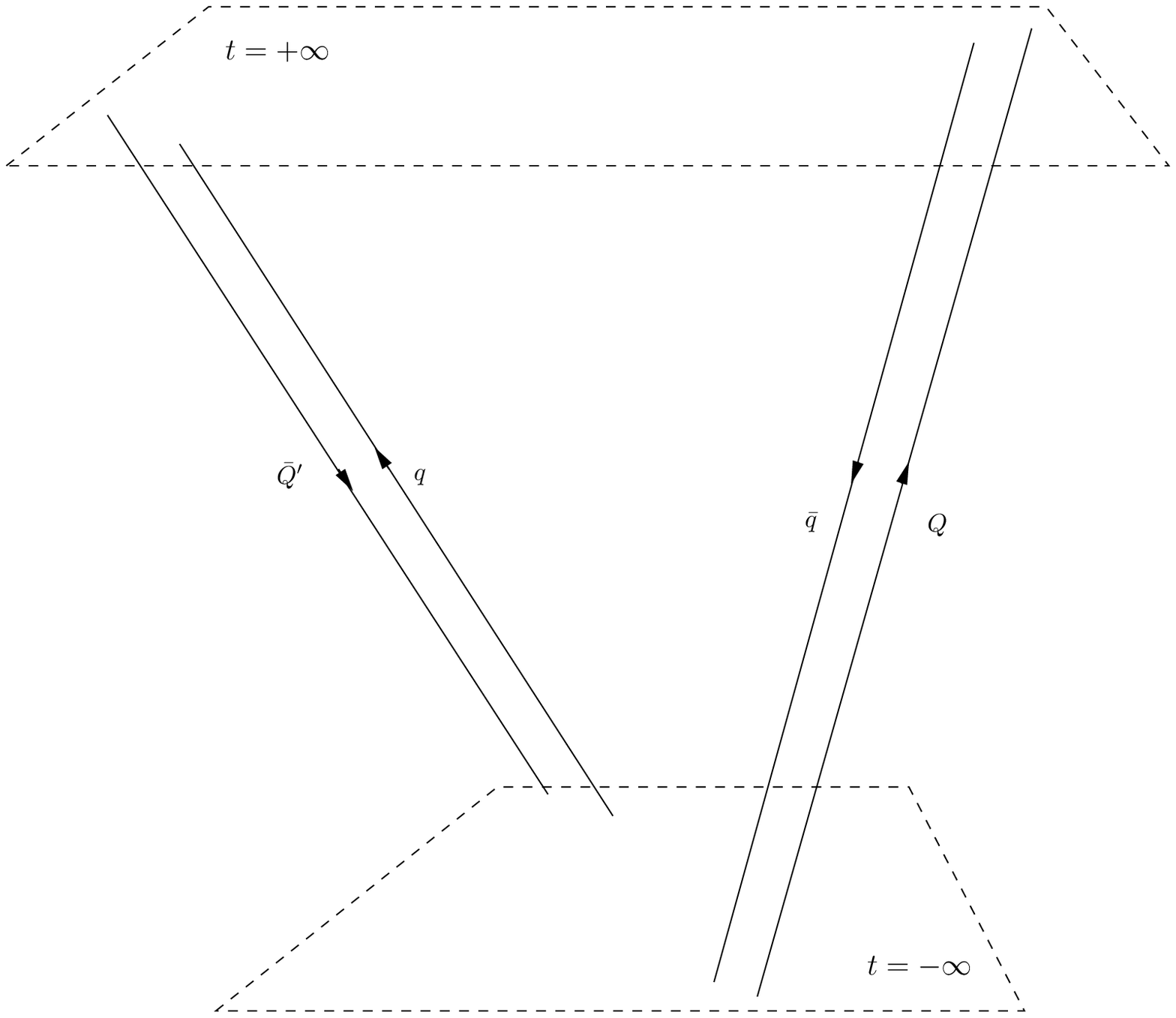}\hspace{0.1cm}
    \includegraphics[width=0.23\textwidth]{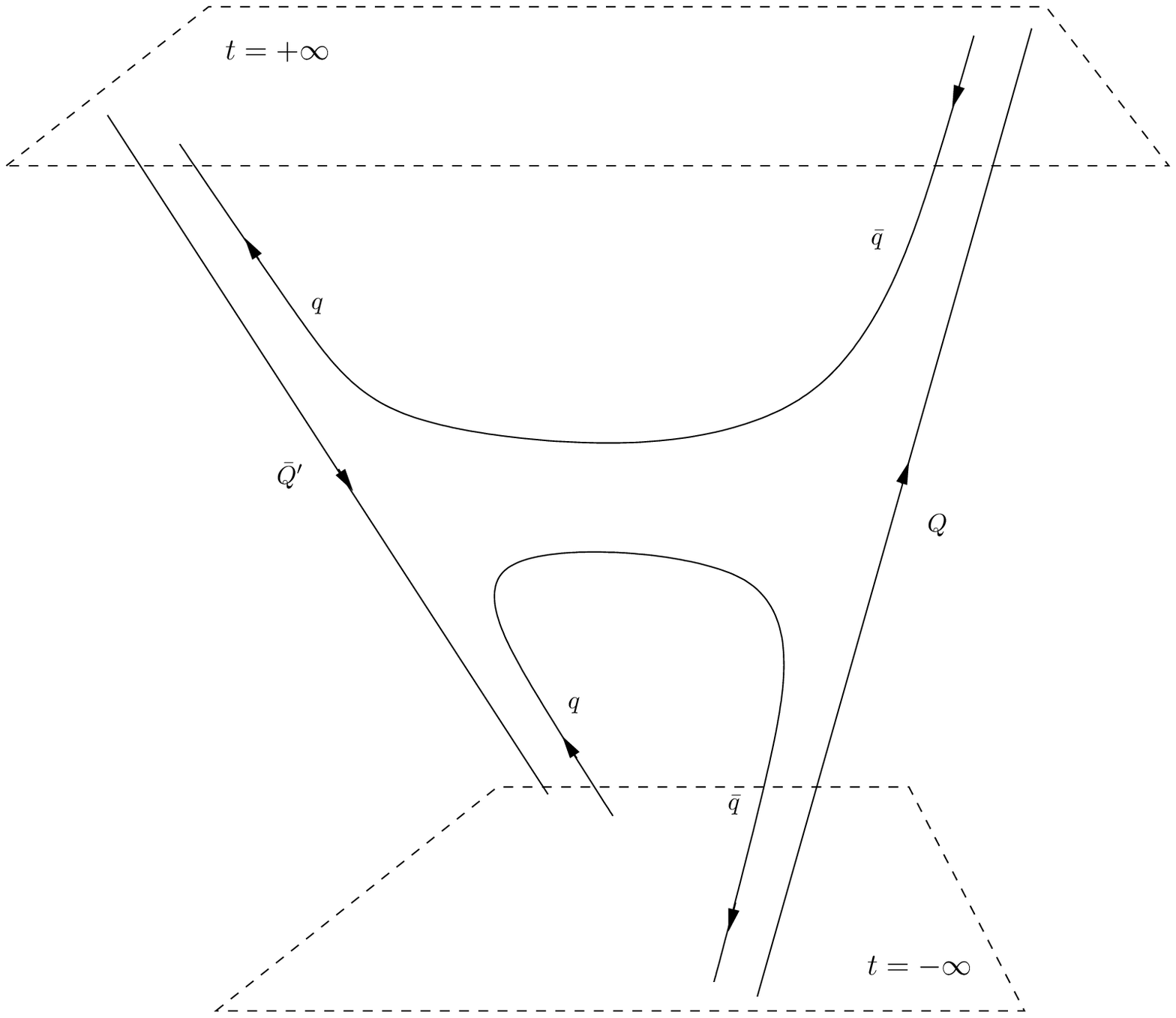}
    \caption{\scriptsize Space-time picture of the
    Pomeron-exchange (PE) process (left) and of the Reggeon-exchange
    (RE) process (right).}
  \label{fig1} 
\end{figure}

The 
starting point is adopting a description of the mesons as wave packets
of transverse colourless dipoles~\cite{Dos}, so reducing the
meson-meson $S$-matrix, $S_{fi}$, to the dipole-dipole ($dd$)
$S$-matrix, $S^{(dd)}_{fi}$,
\begin{equation}
  \label{eq:S_meson_2}
   S_{fi}= \int d\mu\,     S^{(dd)}_{fi}(\mu)\, , \quad
   d\mu=d\mu_1^{\prime *}    d\mu_2^{\prime *} d\mu_1\, d\mu_2 \,, 
\end{equation}
where $\mu$ denotes collectively the various degrees of freedom, and 
the integration measures $d\mu_i^{(*)} $ are defined as 
\begin{equation}
  \label{eq:mes}
  \int d\mu_i^{(*)}  \equiv \int d^2 k_{i\perp} \int_0^1 d\zeta_i
\sum_{s_i,t_i} \psi_{i\,  s_i t_i}^{(*)}(\vec{k}_{i\perp},\zeta_i) \,,
\end{equation}
with $\vec k_{i\perp}$ and $\zeta_i$ the transverse momentum and the 
longitudinal momentum fraction of the quark in meson $i$, $s_i$ and 
$t_i$ the spin indices of the quark and antiquark in meson $i$,
respectively, and $\psi_i$ the wave function for meson $i$.
For later convenience we define also the wave function in coordinate
space, $\wf_{i\,st}(\vr_\perp,\zeta) =  
\sqrt{2\zeta(1-\zeta)2\pi}\int d^2 k_\perp
e^{i\vec{k}_\perp\cdot\vr_\perp}\psi_{i\,st}(\vec{k}_\perp,
\zeta)$, where $\vr_\perp$ is the transverse size of the dipole. One
then performs a  LSZ reduction, 
identifying PE with the parton-elastic process, and
RE with the parton-inelastic one (Fig.~\ref{fig1}):
\begin{equation}
  \label{eq:LSZ}
  S_{fi}^{(dd)}(\mu) = {\cal P}^{(dd)}(\mu) + {\cal R}^{(dd)}(\mu)\,.
\end{equation}
The PE contribution has been investigated in a number of
papers~\cite{Nac,Dos,BN,Meggiola,instantons,Jani1,Jani2,
  adsbounds,lat_pomeron},
and will not be discussed here.  

As regards RE, at this point one exploits the
representation of the fermion propagators as PIs of Wilson
lines running along the trajectories of the partons~\cite{Brandt}, and
the spacetime picture of the process to identify the dominant
contributions to the PI in the large $s$, small $t$  
regime. In the initial stage of the process a ``wee'' (i.e., carrying a
vanishingly small fraction of longitudinal momentum) valence quark $q$
in meson 2, and a ``wee'' valence antiquark $\barq$ in meson 1, enter
the interaction region along the classical straight-line trajectories
of the mesons, then ``bend'' their trajectory, and annihilate
producing gluons; in the final stage of the process, these gluons
produce a ``wee'' $q\barq$ pair, whose components rejoin the
``spectator'' partons to form the mesons in the 
final state.\footnote{Things can go also in the reverse order, with the
production of a fermion-antifermion pair preceeding the
annihilation.} As for the ``spectator'' partons, which carry a
relevant fraction of longitudinal momentum, they travel almost
undisturbed along their eikonal trajectories. This suggests that only
those paths that coincide with the incoming and outgoing eikonal
trajectories of the exchanged partons at early and late times
contribute to the path integral. 

\nin

At this point one has to work out the details, taking care of removing
the internal interactions of the two mesons, that do not take part in
the scattering process. Defining the RE contribution to
elastic meson-meson scattering $\A_{{\cal R}}$, ${\cal R} \!=\! \int
d\mu \,{\cal R}^{(dd)}(\mu)\!=\!
i(2\pi)^4\delta^{(4)}(p_f-p_i)\A_{{\cal R}} $, one obtains the 
following expression for $\A_{{\cal R}}$,
\begin{equation}
  \label{eq:ampli}
  \begin{aligned}
        \A_{{\cal R}}(s,t) = &
    \lim_{\zeta_1\to 1,\,\zeta_2\to 0}  \int d\tilde \mu_{\zeta_1,\zeta_2}
    \A_{{\cal R}}^{(dd)}(s,t;\tilde\mu)\,, 
  \end{aligned}
\end{equation}
where the new integration measure $d\tilde \mu_{\zeta_1,\zeta_2}$ is
\begin{equation}
  \label{eq:tildemu}
  \begin{aligned}
&    \int    d\tilde \mu_{\zeta_1,\zeta_2} =  \int  d^2R_{1\perp}\!\int
    d^2R_{2\perp}\! \int d^2R_{1\perp}'\! \int 
    d^2R_{2\perp}' \\ &\times\sum_{t_1',t_1,s_2',s_2}
    \rho_{1\,t_1't_1}(\vr_{1\perp},\vr_{1\perp}',\zeta_1)
    \rho_{2\,s_2's_2} (\vr_{2\perp},\vr_{2\perp}',\zeta_2)\,, \\
  \end{aligned}
\end{equation}
where $ \rho_{i\,r'r}(\vr_{i\perp},\vr_{i\perp}',\zeta_i)
= \sum_{s}  \wf_{i\,s r'}^*(\vr_{i\perp}',\zeta_i) \wf_{i\,s
  r}(\vr_{i\perp},\zeta_i)$
with $\vr_{i\perp}^{\,(\prime)}$ the dipole size in the initial
(final) state, and $\A_{{\cal R}}^{(dd)}(s,t;\tilde\mu)$ the
RE contribution to $dd$ scattering, 
\begin{equation}
  \label{eq:regge_ampl_dip}
  \begin{aligned}
    &\A_{{\cal R}}^{(dd)}(s,t;\tilde\mu)
\!=\!
-i2s  \f{(2\pi)^2}{m_1 m_2}\f{1}{N_c}
\int\! d^2b_\perp\, e^{i\vec{q}_\perp \cdot\vec{b}_\perp}
\\ & 
\times \int {\cal D C}^{(\wedge)}\int {\cal D
  C}^{(\vee)}\,e^{i4m_qT}e^{-i(m_q-i\epsilon)(L^{(\wedge)} +
  L^{(\vee)})}  
        \\ & \times 
        \Sps_\wedge^{t_1 s_2}[\dot{\cal
          C}^{(\wedge)}\!;\!p_\barq,\!p_q] \Sps_\vee^{s_2'
          t_1'}[\dot{\cal C}^{(\vee)}\!;\!p_q',\!p_\barq'] 
  {\cal U}_{\cal C}[{\cal C}^{(\wedge)},{\cal C}^{(\vee)}]\,.
  \end{aligned}
\end{equation}
Here we have denoted 
${\cal
  C}^{(\wedge),(\vee)}\!=\!(L^{(\wedge),(\vee)},X^{(\wedge),(\vee)})$, $\dot{\cal
  C}^{(\wedge),(\vee)}\!=\!(L^{(\wedge),(\vee)},\dot X^{(\wedge),(\vee)})$,  
with $\int {\cal DC}^{(\wedge),(\vee)}$ defined as
\begin{equation}
  \label{eq:pathdef}
\int {\cal DC}^{(\wedge),(\vee)}=   \int_{2T-L_0}^{2T+L_0} dL^{(\wedge),(\vee)}
\int_{x_i^{(\wedge),(\vee)}}^{x_f^{(\wedge),(\vee)}} \DXtot \,;
\end{equation}
$L^{(\wedge),(\vee)}$ is the length of path $X^{(\wedge),(\vee)}$,
with endpoints\footnote{Note that
  they have been corrected with respect to
  Ref.~\cite{reggeWL}.} 
\begin{equation}
  \label{eq:curved}
  \begin{aligned}
    x^{(\wedge)}_i &= -u_2T + \textstyle\f{R_2}{2} \,, &&&
    x^{(\wedge)}_f &= -u_1T + b - \textstyle\f{R_1}{2} \,, \\ 
    x^{(\vee)}_i &= u_1T + b - R_1' +\textstyle\f{R_1}{2}  \,, &&&
    x^{(\vee)}_f &= u_2T + R_2'-\textstyle\f{R_2}{2} \, ,
  \end{aligned}
\end{equation}
where we have set $b= (0,0,\vec{b}_\perp)$, 
    $R_{1,2}^{(\prime)}=(0,0,\vec{R}_{1,2\perp}^{(\prime)})$. 
In Eq.~\eqref{eq:regge_ampl_dip}, the quantities
\begin{equation}
  \label{eq:sf}
  \begin{aligned}
\Sps_\wedge^{t_1 s_2}
[\dot{\cal C}^{(\wedge)};\!p_\barq,\!p_q] &=
\f{\bar{v}^{t_1}(p_\barq)\Sp_{-T,-T+L}[\dot{X}^{(\wedge)}]
u^{s_2}(p_q)}{2\sqrt{\m_q\m_\barq}}\, 
,\\   
\Sps_\vee^{s_2' t_1'}
[\dot{\cal C}^{(\vee)};\!p_q',\!p_\barq'] &=
\f{\bar{u}^{s_2'}(p_q')\Sp_{-T,-T+L'}[\dot{X}^{(\vee)}]
v^{t_1'}(p_\barq')}{2\sqrt{\m_q'\m_\barq'}}   
\, ,
  \end{aligned}
\end{equation}
are the normalised spin factors, where
\begin{equation}
  \label{eq:defin_2}
  \begin{aligned}
    &\Sp_{\eta,\nu}[\dot{X}] \!=\! \int \DP\, {\cal
      M}_{\eta,\nu}[\dot{X},\!\Pi]\, ,\\ 
    &{\cal M}_{\eta,\nu}[\dot{X},\!\Pi]\!=\!{\rm
      Texp}\left[i\!\int_\eta^\nu\!d\tau 
      \left(\slapi(\tau)\! -\!
        \Pi(\tau)\cdot\dot{X}(\tau)\right)\right]\,  
    ,
  \end{aligned}
\end{equation}
$u^s(p)$ and $v^t(p)$ are bispinors, and
$p_{q}^{(\prime)}=\zeta_2^{(\prime)}p_2$, 
$p_{\barq}^{(\prime)}=(1-\zeta_1^{(\prime)})p_1$. The ``physical''
masses of $q$ and $\barq$ are $\m_q^{(\prime)}=\zeta_2^{(\prime)}m_2$, 
$\m_\barq^{(\prime)}=(1-\zeta_1^{(\prime)})m_1$, while $m_q$ denotes
the bare mass. Finally, the normalised Wilson loop (WL) ${\cal
  U}_{\cal C}$ is 
\begin{equation}
  \label{eq:loop_norm}
 {\cal U}_{\cal C}[{\cal C}^{(\wedge)},{\cal C}^{(\vee)}] \equiv 
\f{\la \W_{\cal
     C}[{\cal C}^{(\wedge)},{\cal C}^{(\vee)}]
     \ra}{
W_{T,|\vec R_{1\perp}|}W_{T,|\vec R_{2\perp}|}W_{T,|\vec R_{1\perp}'|}W_{T,|\vec R_{2\perp}'|}}
\,,
\end{equation}
where $W_{T,|\vec R|}$ is the expectation value of a timelike $T\times
|\vec R|$ rectangular WL, while the WL $\W_{\cal
  C}$ runs along the path ${\cal C}$ made up of ${\cal C}^{(\wedge)}$,
${\cal C}^{(\vee)}$, and the eikonal trajectories $X_Q=u_1\tau + b
+\f{R_1}{2}$ and $X_{\barQ'}=-u_2\tau -\f{R_2}{2}$, $\tau\in[-T,T]$, of
the ``spectator'' quarks, closed by straight-line ``links'' in the 
transverse plane to ensure gauge invariance. The length scale $L_0$
 in Eq.~\eqref{eq:pathdef} is not 
specified for the time being, but it can be fixed from the
experiments; the length scale $T$ is sent to infinity 
at the end of the calculation. 

The $dd$ RE amplitude encodes universal properties of RE, and is
basically described in terms of WLs. However, the limit of
vanishing longitudinal momentum fractions of the exchanged fermions in 
Eq.~\eqref{eq:ampli}, beside ensuring that only ``wee'' partons
contribute to the process in accordance with Feynman's picture of high
energy scattering~\cite{FeynH}, leads to the endpoint behaviour of the
mesonic wave functions affecting the energy dependence of the
amplitude. This is due to the fact that the wave functions are
typically power-like in $\zeta$ near the endpoints, and to the fact
that $\zeta$ has to be taken to zero in the high-energy limit as
$\zeta \sim 1/\sqrt{s}$~\cite{reggeWL}. 

The physical, Minkowskian amplitude in impact-parameter space can be
reconstructed from a corresponding quantity in the Euclidean theory,
i.e., the PI over the Euclidean trajectories of the
exchanged quarks of a Euclidean WL expectation value, defined
essentially as above but with the long sides forming now an angle
$\theta$ in the longitudinal plane. The Minkowskian amplitude is obtained
by means of the AC $\theta\to -i\chi$, $T\to iT$,
$L_0 \to iL_0$~\cite{reggeWL}.

\begin{figure}[t] 
  \centering
    \includegraphics[width=0.32\textwidth]{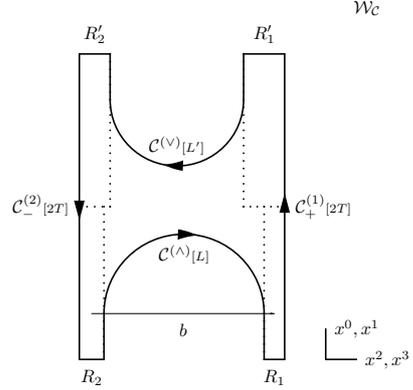}
\caption{\scriptsize Schematic representation of the Wilson loop 
     $\W_{\cal C}$.}
\label{fig2} 
\end{figure} 
\nin

\section{Reggeon Exchange from Gauge/Gravity Duality}
\nin
The (Euclidean) WL formalism is well suited for an
investigation through the gauge/gravity duality~\cite{AdSCFT}. The
idea behind gauge/gravity duality is that strongly coupled gauge
theories can be mapped to gravity theories; in particular, for WL
expectation values the problem is reduced to finding a minimal surface
(MS) in an appropriate metric having the loop contour as
boundary~\cite{Malda}. Although a precise formulation of the duality
is still lacking for confining theories, there is an important feature
that the metric of the dual gravity theory is expected to have, namely
the existence of a characteristic length scale (corresponding to the 
confinement scale) separating the near boundary region with AdS-like
metric from an effectively flat region~\cite{AdSBH}. 
This allows to make the following approximation for the MS relevant to 
WL expectation values~\cite{Jani2}. Near the boundary, the
surface rises almost vertically; the horizon puts a bound on
this rise, and the remaining part of the surface lives in flat
space. This approximation is not sensitive to the details of the
metric, so providing very general results.

Consider now heavy-light mesons of large mass, so that the typical
dipole size is small, and can be neglected in a first
approximation. The MS corresponding to the WL
relevant to RE will be made up of four rectangular
regions, living near the boundary, corresponding to the free
propagation of mesons before and after the interaction, whose
contribution is therefore cancelled out by the normalisation factor;
and of a central strip corresponding to the exchange of a Reggeon. Up
to vertical walls, that lead to the renormalisation of the quark mass,
the strip lives in the flat region, and its geometry is governed by
the heavy quark eikonal trajectories~\cite{Jani,reggeon_duality}. The
most important 
contributions will come therefore from exchanged-quark trajectories
lying on the corresponding helicoid of opening angle $\theta$.
In the saddle-point approximation, one retains only the maximal
contribution, corresponding to the trajectories that minimise the
Euclidean ``effective action'', 
\begin{equation}
  \label{eq:effact}
  S_{{\rm eff,\,E}}  \!\equiv\! \f{1}{2\pi\alpha'_{\rm
        eff}}A + \hat m_q(L^{(\vee)} + L^{(\wedge)}-4T)\,,
\end{equation}
that includes contributions from both
the area $A$ of the surface and the length of its boundaries, subject to
the condition of joining smoothly the incoming and outgoing eikonal
trajectories. An exact solution to this variational problem
exists~\cite{reggeon_duality}, that is independent of $T$ and $L_0$,
but that however is in implicit form and thus not suitable for our
purposes. Since we have to perform the AC, we need to know explicitly
the dependence in $\theta$: this can be achieved making the
approximation of small $\theta$. After going back to Minkowski
space-time, and in the limit of small (renormalised) quark mass $\hat
m_q$, one finds~\cite{reggeon_duality} 
\begin{equation}
  \label{eq:smallm}
  S_{{\rm eff,\,M}} \simeq
  \f{b^2}{4\alpha_{\rm eff}'\chi} - 
  \f{4b\hat m_q}{\chi}  
     + 2\pi^2\alpha_{\rm eff}' \hat m_q^2\,,
\end{equation}
where $S_{{\rm eff,\,M}}\!\equiv\! S_{{\rm eff,\,E}}|_{\theta\to
  -i\chi}$, and $\alpha_{\rm eff}'$ is the effective string tension
corresponding to the confining background.  

In order to obtain the complete expression for the impact-parameter
amplitude in the saddle-point approximation, one should take into
account also spin effects, and quantum fluctuations around the
solution; these terms are however not completely under control at the
moment, and will be neglected in a first approximation, so that the
impact-parameter amplitude takes the Gaussian form $a(\chi,\vec{b})
\approx e^{-S_{{\rm eff,\,M}} }$. One then 
obtains, after Fourier transform, the following $dd$ scattering
amplitude, $\A_{\cal R}^{(dd)}(s,t) 
 \approx
    {\cal T}_0(\chi,t)  + \hat m_q{\cal T}_1(\chi,t)  
+ {\cal O}(\hat m_q^2) $, where\footnote{A factor $(2\alpha_{\rm 
  eff}'\chi)^{-1}$ has been inserted ``by hand'' in order to remove an
extra logarithmic prefactor: this however does not change our
conclusions.} 
\begin{equation}
  \label{eq:smallm2}
    \begin{aligned}
  {\cal T}_0(\chi,t) &=  e^{-\alpha_{\rm eff}'\chi q^2}\,,\\
{\cal T}_1(\chi,t) &=  
\textstyle 8\sqrt{\pi\alpha_{\rm eff}'}\f{\de}{\de\chi}\left[\sqrt{\chi}
  \tilde{I}_0\left(\alpha_{\rm eff}'\chi \f{q^2}{2}\right)  
\right]\, , 
  \end{aligned}
\end{equation}
with $\tilde{I}_0(z) = e^{-z}{I}_0(z)$ ($I_0$: modified Bessel
function). 

Inspecting the singularity structure in the complex-angular-momentum
plane by means of a Mellin transform, one  finds a linear
Regge trajectory: the linearity and the slope are not affected by
prefactors of the type $s^{\delta\alpha}\chi^{n_\chi}b^{n_b}$, that
could come from the neglected terms, and is therefore a robust 
result. In particular, for massless quarks one recovers the result of
Ref.~\cite{Jani}, exactly corresponding to a Regge pole, while for
small but nonzero $\hat m_q$ the singularity contains also a
logarithmic branch point~\cite{reggeon_duality}. Notice that the slope
is equal to the inverse string tension $\alpha_{\rm eff}'$: this
provides a first bridge between the WL formalism and the usual Regge
picture of the exchange between the hadrons of states lying on the
same linear trajectory in the $(J,m^2)$ plane. 
 \section{Conclusions}
 \nin
A formalism based on WLs is now available for the NP study of RE in
SHES. The AC to Euclidean space allows to apply NP techniques to study
this problem. The results obtained to leading order through
gauge/gravity duality are encouraging, and it would be interesting to
compute the effects of quantum fluctuations around the saddle point,
and the effects of dynamical quarks.


\end{document}